\documentclass[useAMS,usenatbib]{mn2e}

\begin{document}

\title[]{Time-Dependent of Accretion Flow with Toroidal Magnetic Field}
\author[A. Khesali and K. Faghei]
{Alireza Khesali \thanks{E-mail: khesali@umz.ac.ir}
and Kazem Faghei \thanks{E-mail: faghei@umz.ac.ir} \\
Department of Physics, Mazandaran University, Babolsar,
Iran \\}
\date{}

\label{firstpage}
\maketitle %

\begin{abstract}
In the present study time evolution of quasi-spherical polytropic
accretion flow with toroidal magnetic field is investigated. The study especially focused
the astrophysically important case in which the adiabatic
exponent $\gamma=5/3$. In this scenario, it was assumed that the angular
momentum transport is due to viscous turbulence and used
$\alpha$-prescription for kinematic coefficient of viscosity. The
equations of accretion flow are solved in a simplified
one-dimensional model that neglects the latitudinal dependence of
the flow. In order to solve the integrated
equations which govern the dynamical behavior of the accretion flow,
self-similar solution was used.
The solution provides some insight into the dynamics of
quasi-spherical accretion flow and avoids many of the strictures of
the steady self-similar solution. The effect of the toroidal
magnetic field is considered with additional variable
$\beta[=p_{mag}/p_{gas}]$, where $p_{mag}$ and $p_{gas}$ are the
magnetic and gas pressure, respectively. The solution indicates a
transonic point in the accretion flow, that this point approaches to
central object by adding strength of the magnetic field. Also, by
adding strength of the magnetic field, the radial-thickness of the
disk decreases and the disk compresses. It was analytically indicated
that the radial velocity is only a function of Alfv'en velocity. The
model implies that the flow has differential rotation and is
sub-Keplerian at all radii.
\end{abstract}


\section{Introduction}
Accretion is the main source of energy in many astrophysical objects
including different types of binary stars, binary X-ray sources,
quasars, and Active Galactic Nuclei (AGN). Though the first
development of accretion theory started a long time ago (Bondi \& Hoyle 1944;
Bondi 1952), intensive development of the theory began
after the discovery of the first X-ray sources (Giacconi et al. 1962)
and quasars (Schmidt 1963). Since the removal of the angular
momentum process operates on slower timescales as compared to
free-fall time, the infalling gas with sufficiently high angular
momentum can form a disklike structure around a central gravitating
body that can be thin or thick depending upon their geometrical
shapes. The models of thin accretion disks are perhaps better
developed and seem to have good observational basis (Shakura \& Sunyaev 1973).
 However, for the thick accretion disks, no fully
developed model exists, and there remain many theoretical
uncertainties about their structure and stability (Banerjee et al. 1995;
Ghanbari \& Abbassi 2004;
Ghanbari et al. 2007).

It is thought that accretion disks, whether in star-forming regions,
in X-ray binaries, in cataclysmic variables, or in the centers of
active galactic nuclei, are likely to be threaded by magnetic
fields. Consequently, the role of magnetic fields has been analyzed
in detail by a number of investigators (Blandford \& Znajek 1977;
Lubow et al. 1994; Banerjee et al. 1995; Shadmehri 2004).A mechanism
 for angular momentum transport is another key ingredient in theory
 of accretion processes and still many theoretical uncertainties
 remain about its nature. As originally pointed out in Lynden-Bell
 (1969) and Shakura \& Sunyaev (1973), a magnetic field can also
contribute to the angular momentum transport. A robust mechanism
of the excitation of magnetohydrodynamical (MHD) turbulence
was shown to operate in accretion disks due to the Magneto-Rotational
Instability (MRI) (Balbus \& Hawley 1998; Machida et al. 1999;
Begelman \& Pringle 2007).

The toroidal magnetic fields have been observed in the outer regions
of YSO discs ((Aitken et al. 1993; Wright et al. 1993; Greaves et al.
1997) and in the Galactic center (Novak et al. 2003; Chuss et al.
2003). Accretion disks containing toroidal magnetic field have
been studied
  by several authors (Fukue \& Okada 1990; Geroyannis \& Sidiras 1992,1993,1995;
   Banerjee et al. 1995; Terquem \& Papaloizou 1996; Machida et al. 1999;
   Liffman \& Bardou 1999;
  Rempel 2006; Begelman \& Pringle 2007;
  Akizuki and Fukue 2006).
   Fukue and Okada (1990) examined the oscillations of
  a gaseous disk which were penetrated by toroidal magnetic fields.
  Geroyannis and Sidiras (1992,1993) described differentially rotating
  polytropic models distorted by toroidal magnetic fields. Also,
  Geroyannis and Sidiras (1995) considered dissipative effects by viscous friction
  of differentially rotating visco-polytropic models that were further distorted by
  a toroidal magnetic field.
   Banerjee et al. (1995) presented a toroidal magnetic field that was generated by
   interaction rotating plasma
  and dipolar magnetic field of central object; they showed that
  toroidal magnetic field has an important effect in structure of the disk.
  Terquem and Papaloizou (1996) studied the a linear stability of a differentially
  rotating disk containing a purely toroidal magnetic field. They presented
  disks containing a purely toroidal magnetic field are always found to be unstable.
  Machida et al. (1999) considered three-dimensional
  global magnetohydrodynamical simulation of a torus treated by
  toroidal magnetic fields.
  Akizuki and Fukue (2006, hereafter AF) examined the effect of toroidal magnetic field on a viscous gaseous disk around a central object under an advection dominated stage. Assuming steady and axisymmetric flow and using steady self-similar method, they presented the nature of the disk was significantly different from that of the weakly magnetized case.\\
  In this study, we want to explore how the dynamic of a rotating and accreting viscous gas
  depends on its toroidal magnetic field. By solving MHD equations for
  accreting gases that are self-similar in time, we will answer this question.
  We assume that turbulent viscosity is due to angular momentum transport of the fluid
  and there is efficient radiation cooling in the flow. This paper is organized as follows. In section 2, the general problem of constructing a model for quasi-spherical magnetized polytropic accretion flow is defined. In section 3, self-similar method for solving the integrated equations which govern the dynamical behavior of the accreting gas is utilized.
The summary of the model is presented in section 4.

\section{General Formulation}

We use spherical coordinate $(r,\theta,\varphi)$ centered on the
accreting object and make the
 following standard assumptions:
\begin{itemize}
  \item [(i)] The accreting gas is a highly ionized gas with infinitive conductivity;
  \item [(ii)] The magnetic field has only an azimuthal component;
  \item [(iii)]The gravitational force on a fluid element is characterized by the Newtonian potential
             of a point mass,$\Psi=-\frac{GM_{*}}{r}$,
  with $G$ representing the gravitational constant and $M_{*}$ standing for the mass of the central star;
  \item [(iv)] The equations written in spherical coordinates are considered in the equatorial
  plane $\theta=\frac{\pi}{2}$ and  terms with any $\theta$ and $\varphi$ dependence are neglected, hence
  all quantities will be expressed in terms of spherical radius $r$ and time $t$;
  \item [(v)] For the sake of simplicity, the self-gravity and general relativistic effects have been neglected;
  \item [(vi)] The equation of state for the accreting gas is
      $p_{gas}=K\rho^{\gamma}$
   with $\gamma$ and $K$ being constant.
\end{itemize}
The macroscopic behavior of such system can be analyzed by
perfect magnetohydrodynamics approximation. As stated in the
introduction, the study focused on analyzing the role of toroidal
magnetic field and viscosity in an accreting gas. Thus, the basic
equations are the continuity Equation,
\begin{equation}\label{a1}
\frac{\partial\rho}{\partial t}
+\frac{1}{r^{2}}\frac{\partial}{\partial r}(r^{2}\rho v_{r})=0,
\end{equation}
the equations of motion,
\begin{equation}\label{a2}
\frac{\partial v_{r}}{\partial t}+v_{r}\frac{\partial
v_{r}}{\partial r}
  +\frac{1}{\rho}\frac{\partial p}{\partial r}
  +\frac{GM_{*}}{r^{2}}
   =r\Omega^{2}-\frac{B_{\varphi}}
     {4 \pi r\rho}\frac{\partial }{\partial r}(rB_{\varphi}),
\end{equation}
\begin{equation}\label{a3}
\rho\left[\frac{\partial}{\partial
t}(r^{2}\Omega)+v_{r}\frac{\partial}{\partial
r}(r^{2}\Omega)\right]=
  \frac{1}{r^{2}}\frac{\partial}{\partial r}\left[\nu\rho r^{4}\frac{\partial \Omega}{\partial r}\right],
\end{equation}
the polytropic equation,
\begin{equation}\label{a4}
p_{gas}=K\rho^{\gamma},
  \end{equation}
and the field freezing equation
\begin{eqnarray}\label{a5}
\frac{\partial B_{\varphi}}{\partial t} +\frac{1}{r}\frac{\partial
}{\partial r}(rv_{r}B_{\varphi})=0,
\end{eqnarray}
where $\Omega(=v_{\phi}/r)$ is the angular speed and
$\nu$ is the kinematic viscosity coefficient.
 As was mentioned
above,
 our understanding of turbulent viscosity is incomplete,
 and for this reason we adopt an empirical prescription,
\ so we employ the usual $\alpha$-prescription (Shakura \& Sunyaev 1973)
 for the viscosity
which we write in the following form for the kinematic coefficient
of viscosity,
\begin{equation}
   \nu = \alpha \frac{p_{gas}}{\rho\Omega_{K}}
\end{equation}
(Narayan \& Yi 1994) where $\alpha$ is constant (Tout 2000; King et. al. 2007)
and $\Omega_{K}$ is the Keplerian angular velocity, Keplerian angular
velocity is defined by
\begin{eqnarray}\label{a6}
\Omega_{K}=\sqrt{\frac{GM_{*}}{r^{3}}}.
\end{eqnarray}
Note that $\nu$ is a function of position and time, since
$\Omega_{K}$ depends on $r$, and $\rho$ varying by $r$ and $t$. To study
the effect of viscosity, $\alpha$ is used as a free parameter.

Before initiating to solve the equations (1)-(5), it is convenient
to non-dimensionalize the equations.
So, the dimensionless variables are introduced according to
\begin{eqnarray}\label{a7}
   \nonumber r\rightarrow\hat{r} r,~~\ t \rightarrow \hat{t}t,~~
    \ \rho\rightarrow\hat{\rho} \rho,~~\ p_{gas} \rightarrow \hat{p}p_{gas},\\
    v_{r}\rightarrow \hat{v} v_{r},~~\ 
\Omega\rightarrow \hat{\Omega}\Omega,~~\
   B_{\varphi}\rightarrow\hat{B} B_{\varphi},
\end{eqnarray}
where
\begin{equation}\label{a8}
\hat{v}=\sqrt{\frac{GM_{*}}{\hat{r}}}=\frac{\hat{r}}{\hat{t}}
=\hat{r}\hat{\Omega},~\
 \hat{p}=\frac{\hat{B}^{2}}{8\pi}={\hat{\rho}\hat{v}^2},
 ~\ K=\frac{GM_{*}}{\hat{r}\hat{\rho}^{\gamma-1}}.
 \end{equation}

Under these transformations and with the use of equations (4),(6),
and (7), equations (1) and (5) do not change, but equations (2) and
(3) become
 \begin{equation}\label{a9}
\frac{\partial v_{r}}{\partial t}+v_{r}\frac{\partial
v_{r}}{\partial r}
  +\gamma{\rho^{\gamma-2}}\frac{\partial\rho}{\partial r}
  +\frac{1}{r^{2}}
   =\frac{v_{\varphi}^{2}}{r}-\frac{2 B_{\varphi}}
     {r\rho}\frac{\partial }{\partial r}(rB_{\varphi}),
\end{equation}
\begin{equation}\label{a10}
  \rho\left[\frac{\partial}{\partial t}(r^{2}\Omega)+v_{r}\frac{\partial}{\partial r}(r^{2}\Omega)\right]=
  \frac{\alpha}{r^{2}}\frac{\partial}{\partial r}\left[  r^{11/2}\rho^{\gamma-1}\frac{\partial\Omega}{\partial r}\right],
\end{equation}

\section{Self-Similar Solutions}
\subsection{Analysis}
To grasp the physics of the accreting viscous gas in a
toroidal magnetic field,
the technique of self-similar analysis proves to be useful.
 Of course, this method is familiar from its wide range of applications in
the full set of equations of MHD in many research fields
of astrophysics.
 In self-similar formulation, the various physical quantities are
expressed as dimensionless functions of a similarity variable, so it
lends itself to a set of partial differential equations, such as
those mentioned above, to be transformed into a set of ordinary
differential equations. A similarity solution, although constituting
only a limited part of problem, is often useful in understanding the
basic behavior of the system. So, in order to seek similarity
solutions for the above equations, a similarity variable $\eta$ is
introduced as
\begin{equation}\label{a11}
\eta=\frac{r}{t^{n}}
\end{equation}
and it is assumed that each physical quantity is given by the following
form:
\begin{equation}\label{a12}
\rho(r,t)=t^{\epsilon_{1}}R(\eta)
\end{equation}
\begin{equation}\label{a13}
v_{r}(r,t)=t^{\epsilon_{2}}V(\eta)
\end{equation}
\begin{equation}\label{a14}
\Omega(r,t)=t^{\epsilon_{3}}\omega(\eta)
\end{equation}
\begin{equation}\label{a15}
B_{\varphi}(r,t)=t^{\epsilon_{4}}B(\eta)
\end{equation}

the exponents
$n,~\epsilon_{1},~\epsilon_{2},~\epsilon_{3},$ and $\epsilon_{4}$
are constant which must be determined. By substituting the
equations (12)-(16) into equations (1), (5),
(10) and (11), the following general results are obtained:
\begin{equation}\label{a16}
\epsilon_{1}=-\frac{2}{3(\gamma-1)},~\
\epsilon_{2}=-\frac{1}{3},~\ \epsilon_{3}=-1,
 ~\ \epsilon_{4}=-\frac{\gamma}{3(\gamma-1)},
\end{equation}
and
\begin{equation}\label{a17}
n=\frac{2}{3}.
\end{equation}
The above results imply each physical quantity retain a similar spacial
shape as the flow evolves, but the radius of the flow
increases proportionally to $t^{2/3}$ .
Also time-dependent density, the pressure and
the toroidal magnetic field are varying by $\gamma$, on the other
hand, they are decreasing by time for $\gamma > 1$.

Here, let us seek time-dependent self-similar of
 the mass accretion rate
\begin{eqnarray}
\dot{M}&=&-4\pi r^{2}\rho v_{r}.
\end{eqnarray}
We can non-dimensionalize the equation (19) under transformation (8) and

\begin{equation}\label{a18}
    \dot{M}\rightarrow\hat{\dot{M}} \dot{M}
 \end{equation}
where
\begin{equation}\label{a19}
    \hat{\dot{M}}=\hat{r}^2\hat{\rho}\hat{v}.
 \end{equation}
Under transformations (8) and (20), equation (19) does not change
and its behavior under similarity quantities that are implied in
equations (12)-(16) can be considered. The similarity solution shows
that the mass accretion rate $\dot{M}$ is proportional to
$t^{(\gamma-5/3)/(\gamma-1)}$. When $\gamma = 5/3$, the mass
accretion rate is independent of time  and decreases in
$1<\gamma<5/3$,
time-dependent behavior of this quantity will be applied in
next section.

Solving equations (1), (10), (11), and (19) under transformations
(12)-(15) in nonmagnetically state, makes it clear that behavior of
physical quantities in the nonmagnetically and the magnetically disk
are the same. The result is one of the strictures of time-dependent
self-similar solution.

Subsequently, the equations (1), (5), (10), and (11) for the
dependence of the physical quantities on the similarity variables
are written as:
\begin{equation}\label{a20}
-\frac{2}{3(\gamma-1)}R+\left(V-\frac{2\eta}{3}\right)\frac{dR}{d\eta}+\frac{R}{\eta^2}
\frac{d}{d\eta}\left(\eta^2V\right)=0,
\end{equation}
\begin{eqnarray}\label{a21}
\nonumber-\frac{V}{3}+\left(V-\frac{2\eta}{3}\right)\frac{dV}{d\eta}
+\gamma
R^{^{\gamma-2}}\frac{dR}{d\eta}+\frac{1}{\eta^{2}}
~~~~~~~~~~~~~~~~\\
=\eta\omega^{2}-\frac{2 B}{\eta R}
\frac{d\left(\eta B\right)}{d\eta},
\end{eqnarray}
\begin{eqnarray}\label{a22}
\nonumber R\left[\frac{1}{3}\left(\eta^{2}\omega\right)+\left(V-\frac{2\eta}{3}
\right)\frac{d}{d\eta}\left(\eta^{2}\omega\right)
\right]~~~~~~~~~~~~~~~~~~~~~\\
=\frac{1}{\eta^{2}}\frac{d}{d\eta}
\left[\eta^{11/2}R^{(\gamma-1)}\frac{d\omega}{d\eta}\right],
\end{eqnarray}
\begin{equation}\label{a23}
-\frac{\gamma}{3(\gamma-1)}B+\left(V-\frac{2\eta}{3}
\right)\frac{dB}{d\eta}+
\frac{B}{\eta}\frac{d}{d\eta}\left(\eta V\right)=0
\end{equation}
This is a system of non-linear ordinary differential equations. Once
$\alpha$ and $\gamma$ are selected, the set of
equations (22)-(25) can be solved. Before solving above equations numerically,
it was found out that equations (22) and (25) imply
\begin{equation}\label{a24}
   V=\frac{2\eta}{3}+C\frac{R}{B^2}
\end{equation}
where $C$ is constant of integration, and will be calculated in
the next section. We can rewrite equation (26) in terms of the
Alfv'en velocity. The Alfv'en velocity in a purely toroidal magnetic
field is $v_{A}^{2}=B_{\varphi}^{2}/4\pi\rho$. By using
transformations of (8), (9), (13), (16), (17), and Alfv'en velocity
equation, equation (26) can be rewritten in the following form
\begin{equation}\label{a25}
   V=\frac{2\eta}{3}+\frac{2C}{A^{2}},
\end{equation}
where $A=v_{A}/\hat{v}$. The result imply that the radial velocity
of a quasi-spherical accretion flow in presence of toroidal magnetic
field is a function of Alfv'en velocity.

\subsection{Inner limit}
When $\gamma=5/3$,
an appropriate asymptotic solution
as $\eta\rightarrow0$ is of the form
\begin{equation}\label{a26}
    R(\eta) \sim R_{0}\eta^{-3/2}
\end{equation}
\begin{equation}\label{a27}
    V(\eta)\sim V_{0}\eta^{-1/2}
\end{equation}
\begin{equation}\label{a28}
    \omega(\eta)\sim \omega_{0}\eta^{-3/2}
\end{equation}
\begin{equation}\label{a29}
    B(\eta)\sim B_{0}\eta^{-1/2}
\end{equation}
in which
\begin{equation}\label{a30}
    R_{0}=\left(\frac{\dot{M}}{12\pi\alpha}\right)^{3/5}
\end{equation}

\begin{equation}\label{a31}
    V_{0}=-3\alpha \left(\frac{\dot{M}}{12\pi\alpha}\right)^{2/5}
\end{equation}

\begin{equation}\label{a32}
    \omega_{0}^{2}=1-\frac{5}{2}\left(\frac{\dot{M}}{12\pi\alpha}\right)^{2/5}
    -\frac{9}{2}\alpha^{2}\left(\frac{\dot{M}}{12\pi\alpha}\right)^{4/5}
\end{equation}

\begin{equation}\label{a33}
    B_{0}^{2}=\beta_{0} \left(\frac{\dot{M}}{12\pi\alpha}\right).
\end{equation}

In order to derive the above relations, the mass accretion rate and
$\beta$ parameter were used, that is ratio of the magnetic pressure
to the gas pressure. When $\eta\rightarrow0$ and $\gamma=5/3$ the
mass accretion rate becomes $\dot{M}\sim -4\pi R_{0} V_{0}$, and the
ratio of the magnetic pressure to the gas pressure becomes
$\beta(\eta)\sim\beta_{0}\eta^{3/2}$, where
$\beta_{0}=B_{0}^{2}/R_{0}^{5/3}$. These relations were applied to
derive equations (32)-(35). In order to present of importance of
magnetic field in the disk, $\beta_{0}$ parameter will be used.

Asymptotic solution shows that $\alpha$ parameter is
 effective in inner edge of the disk and physical quantities
  are sensitive to it, i.e., the radial infall velocity increases by adding $\alpha$,
 the angular velocity
 is sub-Keplerian for all values of $\alpha$, and
 the density and the toroidal magnetic field
 in the inner edge of the disk decrease with increasing $\alpha$.
 These results that are achieved for inner edge of the disk
 are qualitatively consistent with the results of AF. Now, it is possible
to derive approximate constant of integration $C$ in equation (26),
by using equations (28)-(35)
\begin{equation}\label{34}
    C \sim
    -3\alpha\beta_{0}\left(\frac{\dot{M}}{12\pi\alpha}\right)^{4/5}.
\end{equation}

\input{epsf}
\begin{figure}
\centerline{{\epsfxsize=7.0cm\epsffile{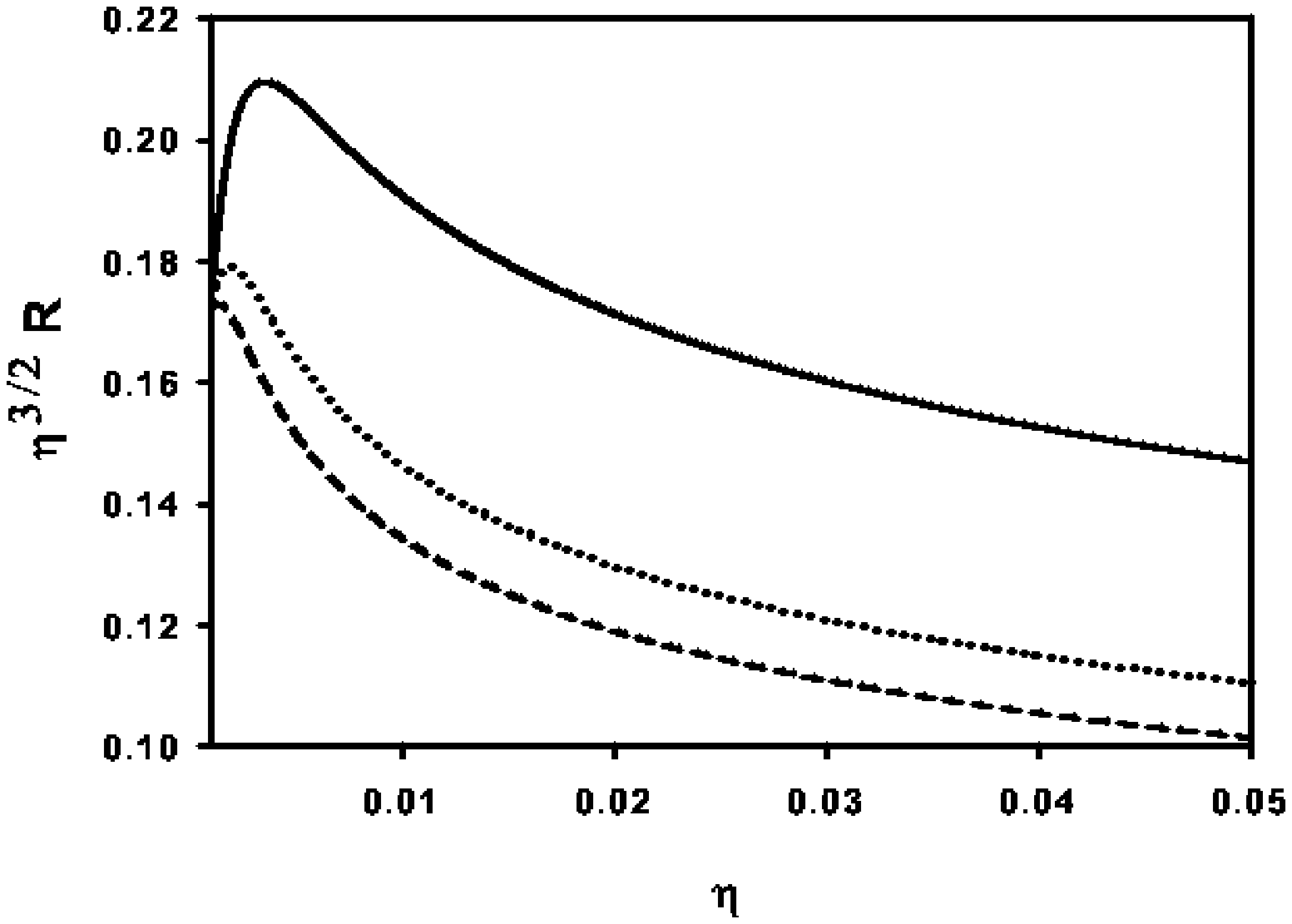}}}
\centerline{{\epsfxsize=7.0cm\epsffile{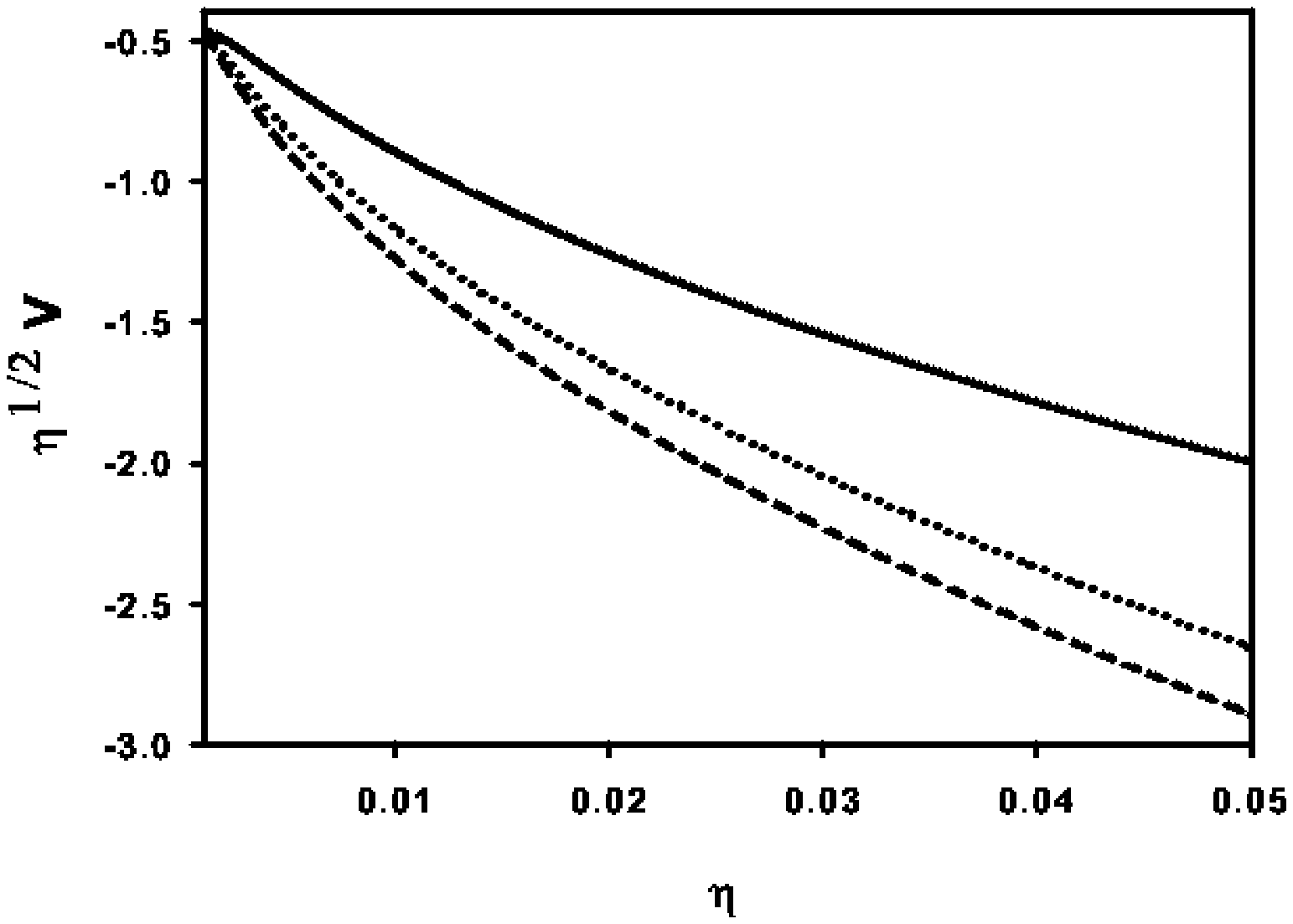}}}
\centerline{{\epsfxsize=7.0cm\epsffile{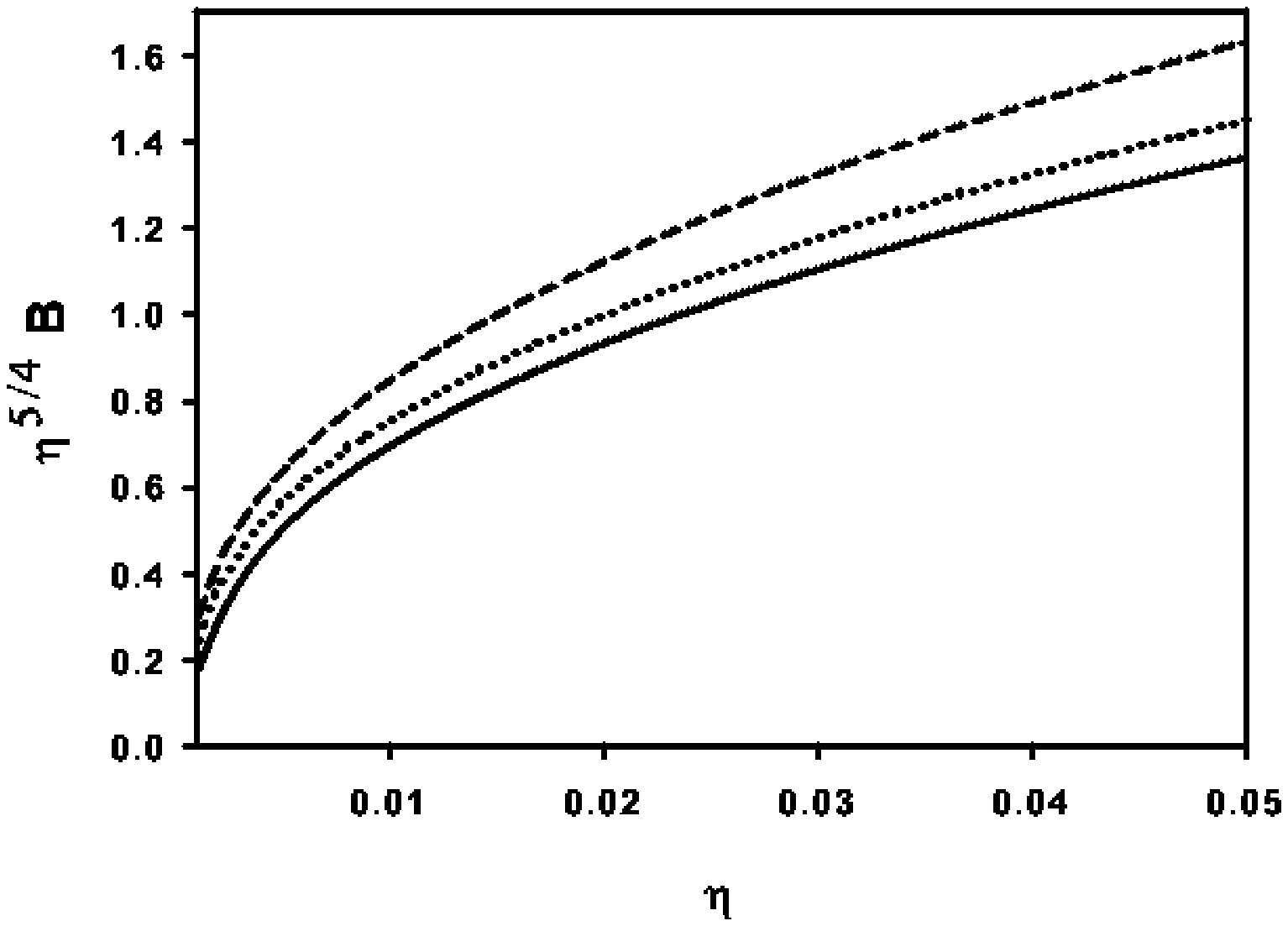}}}
\centerline{{\epsfxsize=7.0cm\epsffile{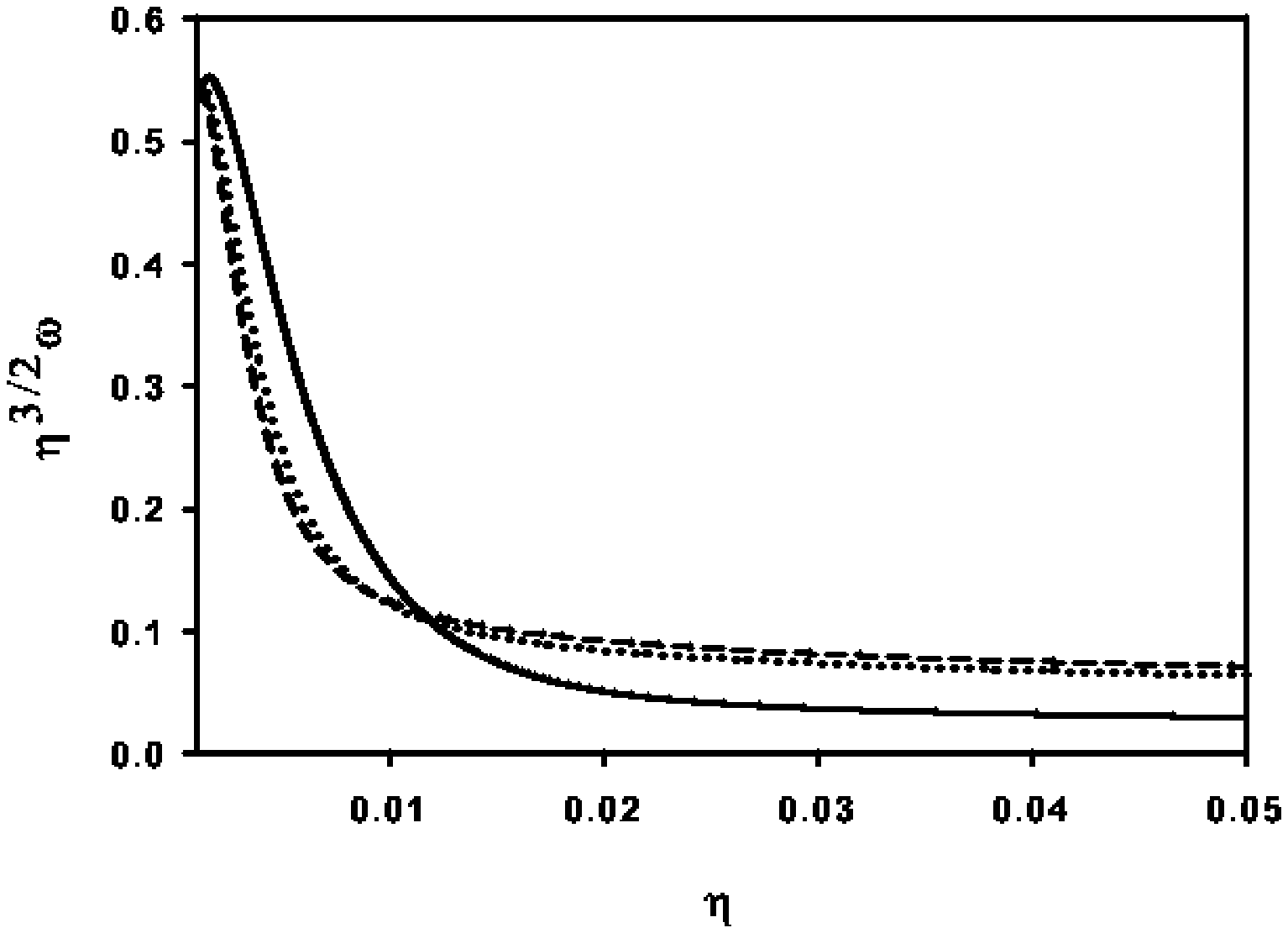}}}
\caption{Time-dependent self-similar solution for $\gamma = 5/3$, $\alpha = 0.5$, and $\dot{M}=1.0$.
The solid lines represent $\beta_{in} = 0.5$, the dotted lines represent
$\beta_{in} = 1.0$, and the dashed lines represent $\beta_{in} = 1.5$ that
$\beta_{in}$ is value of $\beta$ in $\eta_{in}$.}
\end{figure}

\input{epsf}
\begin{figure}
\centerline{{\epsfxsize=7.0cm\epsffile{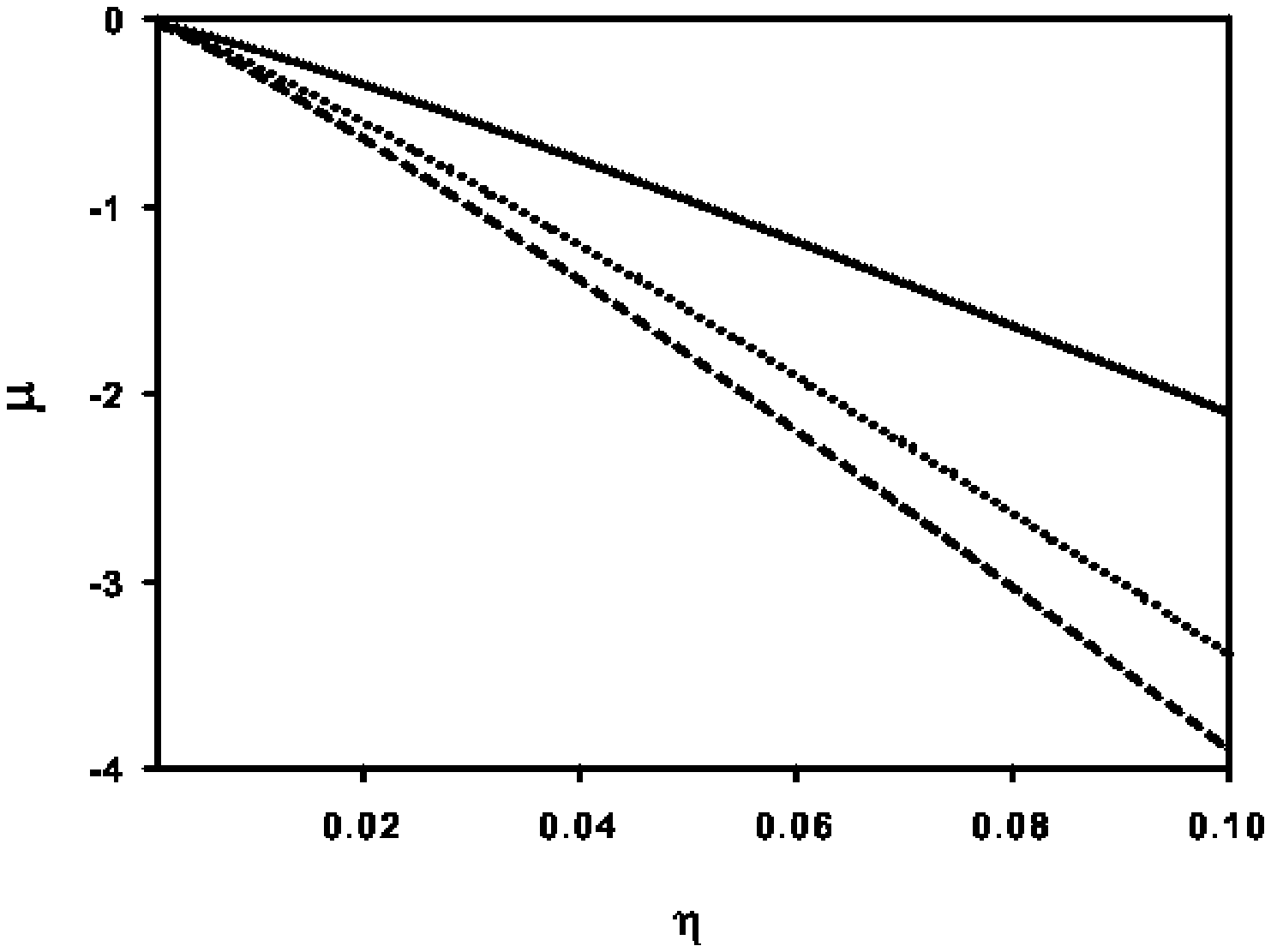}}}
\centerline{{\epsfxsize=7.0cm\epsffile{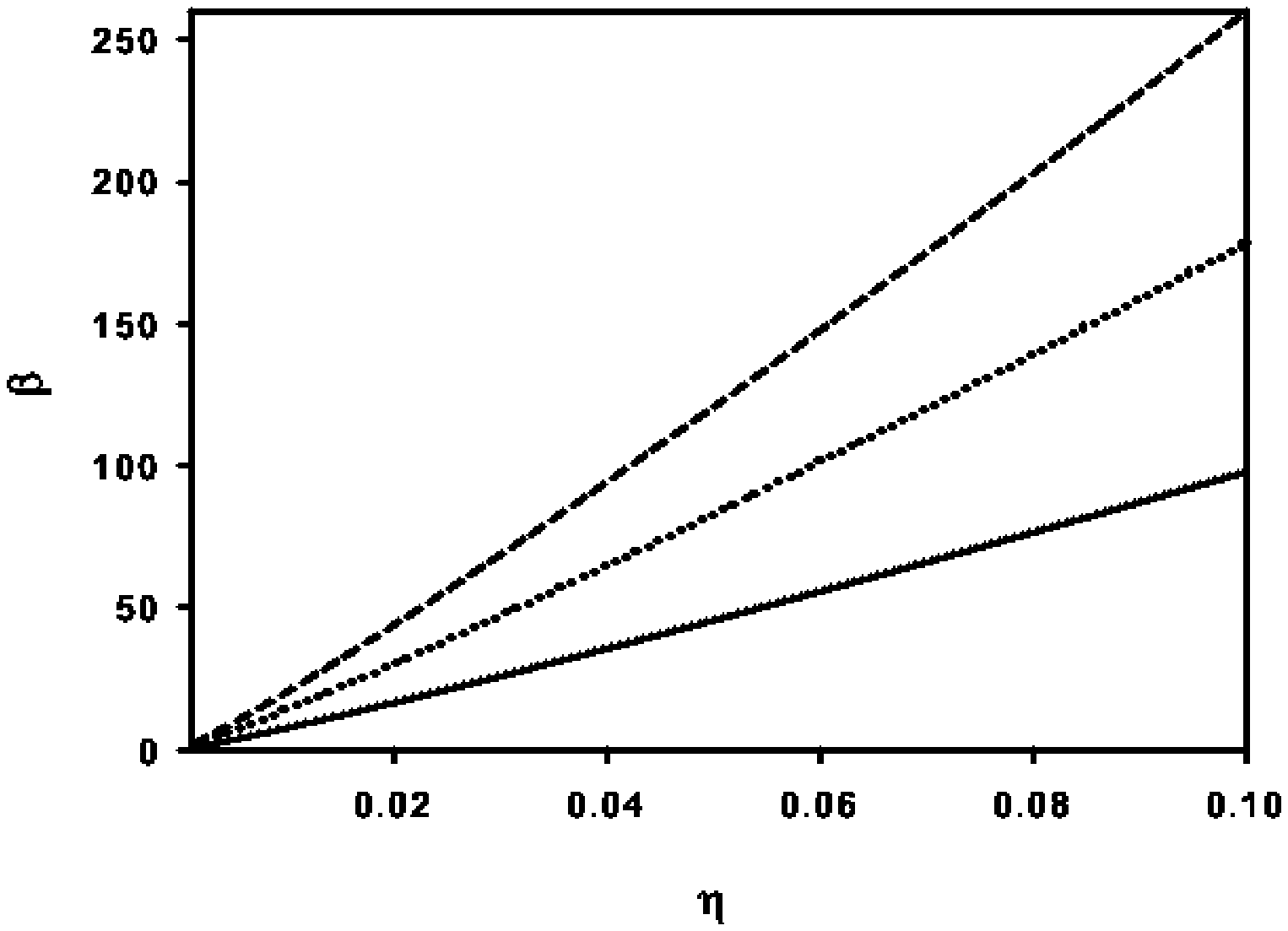}}}
\caption{Time-dependent self-similar solution for $\gamma = 5/3$, $\alpha = 0.5$, and $\dot{M}=1.0$.
The solid lines represent $\beta_{in} = 0.5$, the dotted lines represent
$\beta_{in} = 1.0$, and the dashed lines represent $\beta_{in} = 1.5$ that
$\beta_{in}$ is value of $\beta$ in $\eta_{in}$.}
\end{figure}
\input{epsf}

\subsection{Numerical solution}
If the value of $\eta_{in}$ is guessed, that is a point very near of
the center, the equations can be integrated from this point to
outward through the use of the above expansions. Examples of such
solutions are presented in Figs 1 and 2. The profiles in Fig 1 and 2
are plotted for different $\beta_{in}$ that is amount of $\beta$ in
$\eta_{in}$. From guessed $\eta_{in}$ and $\beta_{in}$, and
equations (28), (31), (32), and (35), we can achieve
 $\beta_{0}=\beta_{in}/\eta_{in}^{3/2}$. The delineated quantities
($\eta^{3/2}R$, $\eta^{1/2}V$, ...) in Figs 1 and 2 are constant at
steady self-similar solutions (Narayan \& Yi 1994; Narayan \& Yi
1995; Shadmehri 2004; Akizuki \& Fukue 2006; Ghanbari et. al. 2007).

By increasing the $\beta$ parameter, which indicates the role of
magnetic field in the dynamics of accretion disks, the
radial-thickness of the disk decreases; the equations (13) and (17)
imply that compression increases by time. Liffman \& Bardou (1999)
and Campbell \& Heptinstall (1998) showed compression of disk in
height direction by effect of toroidal magnetic field, but they did
not consider the effect of toroidal magnetic field in the
radial-thickness of the disk. Also, by adding the $\beta$ parameter,
the radial infall velocity increases; such property is qualitatively
consistent with AF. This is due to the magnetic tension terms, which
dominate the magnetic pressure term in the radial
 momentum equation that assist the radial infall motion.
 The flow is differentially
rotating, although it is highly sub-Keplerian at large radii.

For investigating of existence of transonic point, the
square of the sound velocity is introduced that subsequently can be expressed as
\begin{equation}\label{a35}
v_{s}^2 \equiv \frac{\gamma p_{gas}}{\rho}=\frac{GM}{\hat{r}}t^{-2/3}
\gamma R^{\gamma-1}
\end{equation}
Here, $S=\left(\gamma R^{\gamma-1}\right)^{1/2}$ the \emph{adiabatic sound velocity}
in self-similar flow, which is rescaled in the course of time.
The \emph{Mach number} referred to the reference frame is defined as (Fukue 1984;
Gaffet \& Fukue 1983)
\begin{equation}\label{a36}
    \mu\equiv\frac{v_{r}-v_{F}}{v_{s}}=\frac{V-n\eta}{S}
\end{equation}
where
\begin{equation}\label{a37}
    v_{F}=\frac{dr}{dt}=n\frac{r}{t}
\end{equation}
is the velocity of the reference frame which is moving outward here
as time goes by. The Mach number introduced so far, represents the
\emph{instantaneous} and \emph{local} Mach number of the unsteady
self-similar flow. As we see in Fig 2, there is a transonic point,
that denotes the square of Mach number is equal to unit
($\mu^{2}=1$). By adding strength of the magnetic field, transonic
point approaches to central object. The solution shows that $\beta$
parameter varies by radii and is important at larger radii, while
in steady self similar solution this parameter is constant.
The $\beta$ parameter shows that the dominate pressure in the outer
region of disk is magnetic pressure, that this result is consistent
with observed YSO disks (Aitken et al. 1993; Wright et al. 1993;
Greaves, Holland \& Ward-Thompson 1997).

\section{Summary}

In this paper, the equations of time-dependent quasi-spherical accretion flow with toroidal magnetic field have been solved by semi-analytical similarity methods. The flow is able to radiate efficiency, so we substituted the polytropic equation instead energy equation. A solution was found for the important case $\gamma=5/3$ that has differential rotation and viscous dissipation. The flow avoids many of the strictures of steady self-similar solutions (Narayan \& Yi 1994; Akizuki \& Fukue 2006). Thus, the radial-dependence of calculated physical quantities in this sense are different from steady self-similar solution.

The flow has differential rotation in small radii and has Keplerian
behavior at large radii, that at large radii is similar to steady
self-similar solutions, Also, The flow is sub-Keplerian at all radii
that is consistent with AF when they considered disk in moderate
strength of the magnetic field. The solution shows that in
time-dependent of quasi-spherical accretion flow, there is a
transonic point, where the point approaches to central object by
increasing strength of the toroidal magnetic field. By increasing
strength of the toroidal magnetic field, the radial thickness of the
disk decreases and disk becomes compress.

Here, latitudinal dependence of physical quantities is ignored,
although some authors showed that latitudinal dependence is
important in structure of a disk (Narayan \& Yi 1995; Ghanbari et.
al. 2007). One can investigate latitudinal behavior of such disks.
Also, it is assumed that there is efficient radiation cooling in the
flow and used polytropic equation for energy equation. During recent
years one type of accretion disks has been studied, in which the
energy released through viscous processes in the disk may be trapped
within the accreting gas. This kind of flow is known as
advection-dominated accretion flow (ADAF). Solution of AF shows that
physical quantities of the disk vary by advection parameter. In
future studies, we are going to improve our model with a realistic energy
equation.

\vspace*{1pc}
 We wish to thank the anonymous referee for his/her very constructive comments which helped us to improve the initial version of paper; we would also like to thank  M. Nejad-Asghar,  F. Sohbat
Zadeh, and O. Naser Ghodsi for their helpful discussion.

\end{document}